\documentclass[a4paper,12pt]{article}

\usepackage[latin1]{inputenc}
\usepackage[english]{babel}
\usepackage{amsfonts}
\usepackage{amsmath}
\usepackage{amssymb}
\usepackage{graphicx}
\usepackage{slashed}
\usepackage{latexsym}
\usepackage{amsxtra}
\usepackage{amscd}
\usepackage{amsthm}
\usepackage{dsfont}
\usepackage{appendix}
\usepackage{subfig}
\usepackage{titlesec}
\titleformat{\section}
{\normalfont\large\bfseries}{\thesection}{1em}{}
\titleformat{\subsection}
{\normalfont\normalfont\bfseries}{\thesubsection}{1em}{}

\numberwithin{equation}{section}

\newcommand{\be}{\begin{equation}}
\newcommand{\ee}{\end{equation}}
\newcommand{\ba}{\begin{array}}
\newcommand{\ea}{\end{array}}
\newcommand{\bea}{\begin{eqnarray}}
\newcommand{\eea}{\end{eqnarray}}
\newcommand{\bm}{\begin{matrix}}
\newcommand{\enm}{\end{matrix}}

\renewcommand{\a}{\alpha}
\renewcommand{\b}{\beta}
\newcommand{\g}{\gamma}
\renewcommand{\d}{\delta}
\newcommand{\e}{\varepsilon}
\newcommand{\ve}{\epsilon}

\renewcommand{\l}{\lambda}
\newcommand{\m}{\mu}
\newcommand{\n}{\nu}
\renewcommand{\r}{\rho}
\newcommand{\rb}{\bar{\rho}}

\newcommand{\s}{\sigma}
\renewcommand{\sb}{\bar{\sigma}}
\renewcommand{\t}{\theta}
\newcommand{\tb}{\bar{\theta}}
\newcommand{\p}{\psi}
\newcommand{\pb}{\bar{\psi}}

\renewcommand{\o}{\omega}
\newcommand{\ob}{\bar{\omega}}

\renewcommand{\O}{\Omega}

\newcommand{\ad}{\dot{\alpha}}
\newcommand{\bd}{\dot{\beta}}
\newcommand{\gd}{\dot{\gamma}}
\newcommand{\dd}{\dot{\delta}}

\newcommand{\as}{\underline{a}}
\newcommand{\bs}{\underline{b}}
\newcommand{\cs}{\underline{c}}
\newcommand{\ds}{\underline{d}}

\newcommand{\ks}{\underline{k}}
\newcommand{\ls}{\underline{l}}
\newcommand{\ms}{\underline{m}}
\newcommand{\ns}{\underline{n}}

\newcommand{\G}{\mathcal{G}}
\renewcommand{\H}{\mathcal{H}}

\newcommand{\ph}{\phantom}
\newcommand{\no}{\nonumber}

\def \sba,#1,#2,#3,#4{({\s}^{#1#2})_{#3}^{\phantom #3#4}}
\def \sab,#1,#2,#3,#4{({\sb}\,^{#1#2})^{#3}_{\phantom #3#4}}
\def \sbb,#1,#2,#3,#4{({\s}^{#1#2})_{#3#4}}
\def \saa,#1,#2,#3,#4{({\sb}\,^{#1#2})^{#3#4}}

\def \vba,#1,#2{V_{#1}^{\phantom #1#2}}
\def \va,#1{V^{\,#1}}
\def \vb,#1{V_#1}
\def \obb,#1,#2{{\cal O}_{#1#2}}
\def \oba,#1,#2{{\cal O}_{#1}^{\phantom #1#2}}
\def \oab,#1,#2{{\cal O}^{#1}_{\phantom #1#2}}
\def \oaa,#1,#2{{\cal O}^{#1#2}}

\def \jaa,#1,#2{J^{#1#2}}
\def \jjaa,#1,#2{\overline{J}\,^{#1#2}}
\def \jbb,#1,#2{J_{#1#2}}
\def \jjbb,#1,#2{\overline{J}_{#1#2}}
\def \jab,#1,#2{J^#1_{\ph #1#2}}
\def \jjab,#1,#2{\overline{J}\,^{#1}_{\!\ph #1#2}}
\def \jba,#1,#2{J_{#1}^{\ph #1#2}}
\def \jjba,#1,#2{\overline{J}_{#1}^{\ph #1#2}}

\def \laa,#1,#2{{\l}^{#1#2}}
\def \lab,#1,#2{{\l}^{#1}_{\ph #1 #2}}
\def \lba,#1,#2{{\l}_{#1}^{\ph #1 #2}}
\def \lbb,#1,#2{{\l}_{#1#2}}

\def \waa,#1,#2{w^{#1#2}}
\def \wab,#1,#2{w^{#1}_{\ph #1 #2}}
\def \wba,#1,#2{w_{#1}^{\ph #1 #2}}
\def \wbb,#1,#2{w_{#1#2}}

\def \xaa,#1,#2{X^{#1#2}}
\def \xab,#1,#2{X^{#1}_{\ph #1 #2}}
\def \xba,#1,#2{X_{#1}^{\ph #1 #2}}
\def \xbb,#1,#2{X_{#1#2}}

\def \viel,#1,#2{e_{#1}^{\ph a #2}}
\def \vielp,#1,#2{{e'}_{#1}^{\ph a #2}}
\def \vielt,#1,#2{\widetilde{e}_{#1}^{\ph a #2}}

\def \sviel,#1,#2{E_{#1}^{\ph #1 #2}}

\def \C,#1,#2,#3{C_{#1#2}^{\ph{#1#2}#3}}

\def \fr,#1,#2{\frac{#1}{#2}}
\def \tfr,#1,#2{\tfrac{#1}{#2}}
\def \idz{\int d^2z\,}
\def \idw{\int d^2w\,}
\def \str{\mathrm{STr}}
\def \tr{\mathrm{Tr}}
\def \dst{\displaystyle}
\def \mc{\multicolumn}

\def \ads{AdS$_4\times\mathbb{C}$P$^3$\;}

\def \ads{\mathrm{AdS}_4\times\mathbb{C}\mathrm{P}^3\,}
\def \OSP{\mathrm{OSp}(4|6)}
\def \SO{\mathrm{SO}(3,1)}

\def \sou{so(3,1)\oplus u(3)}
\def \osp{osp(4|6)}

\def \x{\times}
\def \de{\partial}
\def \deb{\bar{\partial}}
\def \onabla{\overline{\nabla}}
\def \oJ{\overline{J}}
\def \oj{\overline{j}}

\def \ln{\mathrm{ln}}
\def \0{|0\rangle}
\def \tg{\widetilde{g}}
\def \tJ{\widetilde{J}}
\def \tj{\widetilde{j}}

\def \ssm{\!\smallsetminus\!}

\begin{document}

\begin{titlepage}
\begin{center}
~\vspace{4cm}

{\Large \textbf {Pure spinor superstring in AdS$_4\x\mathbb{C}$P$^3$\\[1.2ex]with unconstrained ghosts}}

\vspace{2cm}

{\large Marisa Bonini}\footnote{marisa.bonini@fis.unipr.it},
{\large Alessio Camobreco}\footnote{alessio.camobreco@pr.infn.it}

\vspace{.8cm}

\textit{Dipartimento di Fisica ``M. Melloni'', Universit\`a di
Parma}

\& \textit{INFN, Gruppo collegato di Parma}

\textit{Viale G. P. Usberti, 7/A, 43124, Parma, Italy}
\end{center}

\vspace{2cm}

\noindent We construct the action for the pure spinor superstring in
the coset description of $\ads$ superspace, using the variables
which solve the pure spinor condition. As a test of the consistency
of the approach, we use the background field method to verify the
absence of central charge at the second order in the expansion and
to show the one-loop finiteness of the effective action.

\end{titlepage}

\section{Introduction}

The interest in superstring in curved backgrounds has increased
considerably during the last fifteen years as a consequence of the
AdS/CFT correspondence
\cite{maldacena,gubserklebanovpolyakov,wittenads}. Attention first
focused on type IIB $\mathrm{AdS}_5\x\mathrm{S}^5$ superspace, as
main object of the correspondence. The general expression of the
Green-Schwarz superstring in a generic type IIB supergravity
background was known for some time \cite{GHMNT}. In particular, for
the $\mathrm{AdS}_5\x\mathrm{S}^5$ background the explicit form of
the metric and the Wess-Zumino term was found in
\cite{metsaevtseylin}, noting that this superspace is homeomorphic
to the coset supermanifold
${\mathrm{PSU}(2,2|4)}/{\mathrm{SO}(4,1)\x\mathrm{SO}(5)}$ and the
superstring action can be written as a sigma model on this coset.
This approach is the generalization of the flat space construction,
in which the Green-Schwarz superstring is reproduced by a sigma
model on the coset
${\text{SuperPoincar\'e}(\mathrm{D}=10,\mathcal{N}=2)}/{\mathrm{SO}(9,1)}$
\cite{henneauxmezincescu}. The Wess-Zumino term, typical of the
Green-Schwarz action, is given by a 3-form integrated on a three
dimensional volume bounded by the world-sheet. The main property of
the ${\mathrm{PSU}(2,2|4)}/{\mathrm{SO}(4,1)\x\mathrm{SO}(5)}$ coset
is to be a semi-symmetric space, i.e. to admit a
$\mathbb{Z}_4$-grading, and this fact allows to write the
Wess-Zumino term as a world-sheet integral of a 2-form \cite{BBHZZ}.

In general, to quantize the Green-Schwarz action one has to fix the
local fermionic kappa-symmetry. Alternatively, one can introduce
some ghost fields - specifically bosonic spinors - with their
conjugate momenta and provide the action with a BRST symmetry
\cite{berkovits2000}. To assure the on-shell nilpotency of the BRST
charge and the BRST invariance of the action, the ghosts have to
satisfy a peculiar condition and are called pure spinors. (For a
recent review see \cite{oz,bedoyaberkovits,giangreco,mazzucato}.)
The pure spinor approach avoids the presence of the kappa-symmetry.
In particular, in flat space this formalism provides a quadratic
action for the matter fields, hence it does not require to fix the
light-cone gauge and preserves the manifest Poincar\'e covariance.
Solving the pure spinor condition and writing the ghost action in
terms of free fields, it is possible to show \cite{berkovits2000}
the absence of the conformal anomaly and obtain for the Lorentz
currents the same OPE as in the Ramond and Neveu-Schwarz
formulation. Nevertheless the constraint solution breaks the
$\mathrm{SO}(10)$ euclidean Poincar\'e covariance to
$\mathrm{U}(5)$.

Pure spinor superstring naturally extends to curved backgrounds in
supercoset formulation, especially $\mathrm{AdS}_5\x\mathrm{S}^5$
\cite{berkovits2000,berkovitschandia}. In this case the global
Poincar\'e covariance typical of flat space becomes a gauge
covariance under the little group
$\mathrm{SO}(4,1)\x\mathrm{SO}(5)$. To quantize this model one has
to properly take into account, and eventually solve, the pure spinor
constraint resulting from the requirement of BRST invariance.

After the conjecture of Aharony, Bergman, Jafferis and Maldacena
\cite{ABJM} the attention has been extended to type IIA superstring
in the $\ads$ background as dual of a $\mathcal{N}=6$ superconformal
Chern-Simons-matter theory in three dimensions. The bosonic part of
the supercoset $\OSP/\SO\x\mathrm{U}(3)$ is homeomorphic to $\ads$,
hence it is natural to write the Green-Schwarz superstring as a
sigma model on this coset \cite{arutyunovfrolov,stefanski,uvarov},
like in the $\mathrm{AdS}_5\x\mathrm{S}^5$ case. However the
supercoset $\OSP/\SO\x\mathrm{U}(3)$ has 24 fermionic degrees of
freedom instead of 32, thus it does not completely describe the IIA
$\ads$ superspace \cite{gomissorokinwulff}. As discussed in
\cite{arutyunovfrolov}, the sigma model action can be thought as the
Green-Schwarz action with 8 degrees of freedom gauged away by using
half of the kappa-symmetry parameters, indeed the sigma model still
presents a local fermionic invariance of rank 8. Nevertheless, for
particular configurations, such as a string moving only in the AdS
part of the background, the rank of the fermionic symmetry becomes
12 and the coset model does not contain all physical fermionic
degrees of freedom.

The supercoset $\OSP/\SO\x\mathrm{U}(3)$ is a semi-symmetric
superspace and admits pure spinor formulation for the superstring
\cite{fregrassi,bonelligrassisafaai,DFGT} completely analogous to
the $\mathrm{PSU}(2,2|4)/\mathrm{SO}(4,1)\x\mathrm{SO}(5)$
background. The aim of this work is to present a coset formulation
of the superstring in $\ads$ with the ghost action written in terms
of the variables which solve the pure spinor condition. After having
identified the ghost degrees of freedom, we replace the pure spinor
action with a new action for these variables by imposing
$\SO\x\mathrm{U}(3)$ gauge covariance. We use the background field
method to check the consistency of our model. In particular we show
the vanishing of the conformal anomaly up to the second order in the
background expansion parameter and the absence of divergent
contributions in the one-loop effective action.

The paper is organized as follows. In section 2 we resume the main
results for the pure spinor superstring in semi-symmetric spaces and
in section 3 we specify the $\ads$ case as $\OSP/\SO\x\mathrm{U}(3)$
supercoset. In section 4 we give the solution of the pure spinor
constraint in term of independent ghosts and auxiliary variables. In
section 5 we present the action term for these fields and quantize
the model using the background field method. Finally, in section 6,
we make perturbative computations of the central charge and the
beta-function. In the appendices we summarize our conventions and
give the $\OSP$ superalgebra in a form suitable to handle the pure
spinor constraint.

\section{Pure spinor superstring in semi-symmetric superspaces}

Let us consider a superspace described by a supercoset manifold
$G/H$, where the bosonic part of the supergroup $G$ - named
$\mathrm{Bos}[G]$ - gives the isometries of the space and $H$ is its
bosonic stability subgroup. If the Lie superalgebra $\G$ of $G$
admits an automorphism $\O$ involutive on $\mathrm{Bos}[\G]$, the
superspace is said semi-symmetric \cite{serganova}. Thus, defining
$\H_k$ the eigenspace of $\O$ associated with the eigenvalue $i^k$,
$k=0,1,2,3$, $\G$ can be $\mathbb{Z}_4$-graded as \be
\G=\bigoplus_{k=0}^3\mathcal{H}_k\,.\no\ee By definition
$\O\left([A,B]\right)=[\O(A),\O(B)]$ for all $A,B\in\G$, hence if
$H_{k,l}\in\H_{k,l}$ one has
\begin{align*}[H_k,H_l]\in\mathcal{H}_{k+l\,|\,\mathrm{mod}\,4}\end{align*}
i.e. $\H_0$ and $\H_2$ are the bosonic eigenspaces, while $\H_1$ and
$\H_3$ are the fermionic ones. In particular $\H_0$ is a closed
subalgebra and generates the subgroup $H$. Semi-symmetric spaces and
their automophisms have been completely classified in
\cite{serganova}. For the types corresponding to $\mathrm{PSU}(n|n)$
and $\mathrm{OSP}(2n|2n+2)$ symmetries, superstring admits a sigma
model description \cite{zarembo}. In these cases the action is
written in terms of the canonical form $J=g^{-1}dg$, with $g\in G$,
that takes values in $\G$, satisfies the Maurer-Cartan equation \be
dJ+J\wedge J=0\no\ee and decomposes as\be J=\sum_{k=0}^3J_k\,,\qquad
J_k\in\H_k\,.\no\ee $J$ is invariant under global left
multiplication $g\rightarrow g'g$ with $g'\in G$, while under a
local right multiplication $g\rightarrow gh$ with $h\in H$, $J_0$
transforms as a gauge connection \be J_0\rightarrow
h^{-1}J_0h+h^{-1}dh\no\ee and $J_{1,2,3}$ transform according to the
adjoint representation of $h$, i.e. like matter fields \be
J_{1,2,3}\rightarrow h^{-1}J_{1,2,3}h\,.\no\ee In complex worldsheet
coordinates, every canonical form $J_k$ has two components
$J_k,\oJ_k$ transforming as $(1,0)$ and $(0,1)$ worldsheet tensor
respectively.

The pure spinor superstring action
\cite{berkovits2000,berkovitschandia} is given by the sum of a
matter and a ghost part \be\label{purespinoraction}
S_{PS}=S_{matter}+S_\l\ee with \begin{align}
&S_{matter}=\fr,1,{2\pi\a'}\idz\str\!\left[\fr,1,2J_2\overline{J}_2+\fr,3,4J_3\overline{J}_1+\fr,1,4J_1\overline{J}_3\right]\label{Smatter}\end{align}
and\begin{align}
&S_\l=-\fr,1,{2\pi\a'}\idz\str\!\left[w_3\overline{\nabla}\l_1+w_1\nabla\l_3+\{w_3,\l_1\}\{w_1,\l_3\}\right]\,.\label{Slambda}\end{align}
The ghost fields $\l_1$, $\l_3$ are worldsheet scalars and take
values in the fermionic eigenspaces $\H_1$ and $\H_3$ respectively,
while their conjugate momenta $w_3\in\H_3$ and $w_1\in \H_1$ are
holomorphic and anti-holomorphic one-forms. The gauge field $J_0$
only appears in the covariant derivative
\be\label{covariantderivative} \nabla\l\equiv\de\l+[J_0,\l]\no\ee
and couples the ghost to the matter sector.

The BRST transformation acts on the group element $g$ by right
multiplication, $Q(g)=g(\l_1+\l_3)$. From $J=g^{-1}dg$ one
immediately obtains \begin{align*}
&Q(J_{2n})=[J_{2n+3},\l_1]+[J_{2n+1},\l_3]\,,\\[2ex]
&Q(J_{2n+1})=\nabla\l_{2n+1}+[J_2,\l_{2n+3}]\,,\end{align*} where
$n=0,1$ and all indices are modulo 4. For the ghost fields one
assumes \be Q(\l_1)=0\,,\qquad Q(\l_3)=0\,,\qquad
Q(w_3)=J_3\,,\qquad Q(w_1)=\oJ_1\,,\no\ee according to
$\mathbb{Z}_4$-grading and conformal weight. The requirement of BRST
invariance for the pure spinor action yields the
conditions\be\label{ghostconstraint}\{\l_1,\l_1\}=0\,,\qquad
\{\l_3,\l_3\}=0\,,\ee that correspond to the pure spinor constraint
in flat space. Because of (\ref{ghostconstraint}), the action
(\ref{purespinoraction}) has an additional local invariance that
affects the antighost fields only \be\label{anotherlocalinvariance}
\d w_3=[\l_1,\Omega_2]\,,\qquad\d w_1=[\l_3,\Omega_2]\ee with
$\Omega_2\in\mathcal{H}_2$. Moreover the conditions
(\ref{ghostconstraint}) assure $Q^2=0$ on shell up to gauge
transformations \cite{berkovits2004,berkovits2008}.

The explicit form of the pure spinor action (\ref{purespinoraction})
and pure spinor constraint (\ref{ghostconstraint}) depends on the
superspace, i.e. on the supercoset. The
$\mathrm{AdS}_5\x\mathrm{S}^5$ case was studied in \cite{berkovits2000,berkovitschandia}
using the $\mathrm{PSU}(2,2|4)/\mathrm{SO}(4,1)\x\mathrm{SO}(5)$ coset, while
the $\ads$ case was studied in \cite{fregrassi,bonelligrassisafaai,DFGT}
using the $\OSP\,/\,\mathrm{SO}(3,1)\!\x\!\mathrm{U}(3)$ coset.

\section{$\OSP\,/\,\mathrm{SO}(3,1)\!\x\!\mathrm{U}(3)$ supercoset}

Let us introduce \cite{kac} the $(4+6)\x(4+6)$ even supermatrix \be
M=\begin{pmatrix}A&X\\Y&B\end{pmatrix}\no\ee with Grassmann even
entries for $A$, $B$ and Grassmann odd entries for $X$, $Y$. We
define the supertranspose of $M$ \be
M^{st}=\begin{pmatrix}A^t&-Y^t\\X^t&B^t\end{pmatrix}\no\ee and the
$(4|6)$ metric \be \quad
K=\begin{pmatrix}C_4&0\\0&\textbf{1}_6\end{pmatrix}\,,\no\ee where
$C_4$ is a real, antisymmetric matrix with $C_4^{\ph 4
2}=-\mathbf{1}_4$ that can be chosen as the 4-dimensional charge
conjugation matrix (see Appendix \ref{appendixconventions}). By
definition, $M$ is in the superalgebra $\osp$ of the orthosymplectic
supergroup $\OSP$ if \be M^{st}K+KM=0\,,\no\ee i.e. \be
A^tC_4+C_4A=0\,,\qquad\quad B^t+B=0\,,\qquad\quad
Y^t-C_4X=0\,,\no\ee that gives \be
\mathrm{Bos}[\OSP]\cong\mathrm{Sp}(4)\x\mathrm{SO}(6)\cong\mathrm{SO}(3,2)\x\mathrm{SU}(4)\,.\no\ee

There exist two real antisymmetric matrices $K_4$, $K_6$ of order 4
and 6 respectively, with the properties $[K_4,C_4]=0$, $K_4^{\ph 4
2}=-\textbf{1}_4$, $K_6^{\ph 6 2}=-\textbf{1}_6$, so that \be
\Omega(M)\equiv\begin{pmatrix}K_4A^tK_4&K_4Y^tK_6\\-K_6X^tK_4&K_6B^tK_6\end{pmatrix}\no\ee
is an automorphism involutive on $sp(4)\oplus so(6)$ giving the
$\mathbb{Z}_4$-grading of $\osp$. In particular the $\O$-invariant
subalgebra is $\mathcal{H}_0=so(3,1)\oplus u(3)$ and the bosonic
part of the supercoset $\OSP/\mathrm{SO}(3,1)\x\mathrm{U}(3)$ is \be
\mathrm{Bos}\left[\fr,{\OSP},{\mathrm{SO}(3,1)\x\mathrm{U}(3)}\right]
\cong\fr,{\mathrm{SO}(3,2)},{\mathrm{SO}(3,1)}\x\!\fr,{\mathrm{SU}(4)},{\mathrm{U}(3)}\cong\ads\no\ee
as required. It is important to note that
\be\left[\Omega(M)\right]^*=\Omega(M^*)\,,\no\ee thus, for all
$H_3\in\mathcal{H}_3$, $\Omega({H_3}^*)=i{H_3}^*$ i.e.
${H_3}^*\in\mathcal{H}_1$. Similarly, for all $H_1\in\mathcal{H}_1$,
${H_1}^*\in\mathcal{H}_3$ and one can conclude that there is a
one-to-one correspondence between $\mathcal{H}_1$ and
$\mathcal{H}_3$.\footnote{In general this property holds for all
$\mathrm{OSp}$ semi-symmetric spaces and for most of the
$\mathrm{PSU}$ ones. In particular it holds for
$\mathrm{PSU}(2,2|4)$ \cite{zarembo}.}
\\
\\
We choose the
$\mathrm{SO}(3,1)\x\mathrm{U}(3)\x\mbox{``translations''}$ basis for
the $\OSP$ superalgebra (see Appendix \ref{appendixalgebra}): The
bosonic generators are
\bea   &&\H_0:\left\{M^{mn}\in so(3,1)\,,\quad\vba,a,b\in u(3)\right\}\,,\no\\[2ex]
       &&\H_2:\left\{P^m\in so(3,2)\ssm so(3,1)\,,\quad V_a\,,V^a\in su(4)\ssm u(3)\right\}\no\eea
and the fermionic generators are\be
\H_1:\left\{\obb,\a,a\,,\;\oaa,\ad,a\right\}\,,\qquad\qquad
    \H_3:\left\{\oba,\a,a\,,\;\oab,\ad,a\right\}\no\ee
with \be m,n=0,1,2,3\qquad a,b=1,2,3\qquad \a,\ad=1,2\,.\no\ee
$M^{mn}$ and $P^m$ generate the rotations and the translations in
$\mathrm{AdS}_4$ respectively, while $\vba,a,b$ and $V_a\,,V^a$ play
analogous role in $\mathbb{C}\mathrm{P}^3$. The complete expression
for $\OSP$ superalgebra is given in Appendix \ref{appendixalgebra}.
The non-vanishing supertraces are \begin{align}\label{supertraces}
&\str(M_{kl}M_{mn})=\eta_{k[m}\eta_{n]l}\,,&&\str(P_m
P_n)=\eta_{mn}\,,\no\\[1ex]
&\str(\vba,a,b\vba,c,d)=-2\d_a^{\ph ad}\d_c^{\ph cb}\,,&&\str(\vb,a\va,b)=-\d_a^{\ph
ab}\,,\\[.5ex]
&\str(\obb,\a,a\oba,\b,b)=i\e_{\a\b}\d_a^{\ph ab}\,,
                                                &&\str(\oaa,\ad,a\oab,\bd,b)=i\e^{\ad\bd}\d_b^{\ph
ba}\,,\no\end{align} with $\eta_{mn}=\mathrm{diag}(+,-,-,-)$,
$\eta_{k[m}\eta_{n]k}=\eta_{km}\eta_{ln}-\eta_{kn}\eta_{lm}$ and
${\e}^{12}=-{\e}^{21}=-{\e}_{12}={\e}_{21}=1$. Thus we can define
the components of the matter fields
\begin{align*}
&J_0=J^{mn}M_{mn}+\jab,a,b\vba,a,b\,,&&J_2=J^mP_m+J^aV_a+J_aV^a\,,\\[1ex]
&J_1=\jaa,\a,a\obb,\a,a+\jbb,\ad,a\oaa,\ad,a\,,&&J_3=\jab,\a,a\oba,\a,a+\jba,\ad,a\oab,\ad,a
\end{align*} and of the ghost/antighost fields \begin{align*}
&\l_1=\laa,\a,a\obb,\a,a+\lbb,\ad,a\oaa,\ad,a\,,&&\l_3=\lab,\a,a\oba,\a,a+\lba,\ad,a\oab,\ad,a\,,
\\[1ex]
&w_3=\wab,\a,a\oba,\a,a+\wba,\ad,a\oab,\ad,a\,,&&w_1=\waa,\a,a\obb,\a,a+\wbb,\ad,a\oaa,\ad,a\,.
\end{align*} Using the supertraces (\ref{supertraces}) we can write the pure spinor action
(\ref{purespinoraction}) for the $\ads$ background explicitly. The
matter term (\ref{Smatter}) becomes
\begin{align}\label{explicitSmatter}
&S_{matter}=\fr,R^2,{2\pi}\idz\left[\fr,1,2\eta_{mn}J^m\overline{J}^n-\fr,1,2J_a\overline{J}^a-\fr,1,2J^a\overline{J}_a\right.\no\\
&\ph{S_{matter}=}\left.-\fr,i,4\e_{\a
               \b}\left(3\jab,\a,a\jjaa,\b,a+\jaa,\a,a\jjab,\b,a\right) -
               \fr,i,4\e^{\ad\bd} \left(3{\jba,\ad,a} {\jjbb,\bd,a}+{\jbb,\ad,a}
               {\jjba,\bd,a} \right)\right]\end{align}
and ghost term (\ref{Slambda}) becomes
\begin{align}\label{explicitSlambda}
&S_{\l}=\fr,R^2,{2\pi}\idz\left[-i\e_{\a\b}\left(w^\a_{\ph\a a}\overline{\nabla}\l^{\b a} + w^{\a a}\nabla\l^{\b}_{\ph\b a}\right)
                -i\e^{\ad\bd}\left(w_{\ad}^{\ph \ad a}\overline{\nabla}\l_{\bd a}+w_{\ad a}\nabla\l_{\bd}^{\ph \bd a}\right)\right.\no\\
&\ph{S_{\l}}+\fr,1,{8}\eta_{km}\eta_{ln}\left(\wab,\a,a\sbb,k,l,\a,\b\laa,\b,a+
        \wba,\ad,a\saa,k,l,\ad,\bd\lbb,\bd,a\right)\!\left(\waa,\g,b\sbb,m,n,\g,\d\lab,\d,b+\wbb,\gd,b\saa,m,n,\gd,\dd
        \lba,\dd,b\right)\no\\
&\ph{S_{\l}}-\left.\fr,1,2\left(\e_{\a\b}\wab,\a,b\laa,\b,a-\e^{\ad\bd}\wba,\ad,a\lbb,\bd,b\right)\!
        \left(\e_{\a\b}\waa,\a,b\lab,\b,a-\e^{\ad\bd}\wbb,\ad,a\lba,\bd,b\right)\right]\,,
\end{align}
where the coupling constant is given naturally by the
$\mathrm{AdS}_4$ radius $\a'\equiv 1/R^2$.

\section{Solution of the constrain}

By means of the $\OSP$ superalgebra the ghost constraints
(\ref{ghostconstraint}) become

\be\label{explicitconstraint} \ba{l}
\ve_{abc}\laa,\a,a\e_{\a\b}\laa,\b,b=0\,,\\[1.5ex]
\ve^{abc}\lbb,\ad,a\e^{\ad\bd}\lbb,\bd,b=0\,,\\[1.5ex]
\laa,\a,a (\s^m)_\a^{\ph \a \bd}\lbb,\bd,a=0\,, \ea\qquad\qquad
\ba{l}
\ve^{abc}\lab,\a,a\e_{\a\b}\lab,\b,b=0\,,\\[1.5ex]
\ve_{abc}\lba,\ad,a\e^{\ad\bd}\lba,\bd,b=0\,,\\[1.5ex]
\lba,\ad,a(\overline{\s}^m)^{\ad}_{\ph \ad \b}\lab,\b,a=0\,.\ea\ee
\\
The constraint on $\l_1$ can be solved setting
\cite{fregrassi}\be\label{solutionlambda1} \laa,\a,a=\t^\a u^a\,,
\qquad\qquad \lbb,\ad,a=\p_{\ad} v_a\ee with the condition
\be\label{uv}u^a v_a=0\,.\ee Moreover, exploiting the possibility of
rescaling the $\t$ and $\p$ variables, one can normalize $u$ and $v$
and set \be\label{norm} |u|^2\equiv u_a^*u^a=1\,,\qquad |v|^2\equiv
{v^a}^*v_a=1\,.\ee In this way the constraint on $\l_1$ - i.e. the
first column of (\ref{explicitconstraint}) - translates into the
conditions (\ref{uv}) and (\ref{norm}).

The ghost fields $\t^\a$ and $\p_{\ad}$ are unconstrained spinors
which transform under the $(\mathbf{2},\mathbf{1})$ and
$(\mathbf{1},\mathbf{2})$ representations of $\SO$ respectively
\be\label{thetatransformation}
\d\t^\a=\fr,1,4\t^\b(\xi_{mn}\s^{mn})_{\b}^{\ph\b\a}\,,
\qquad\quad\d\p_{\ad}=\fr,1,4\p_{\bd}(\xi_{mn}{\sb}^{mn})^{\bd}_{\ph\bd\ad}\,,\no\ee
with $\xi_{nm}=-\xi_{mn}$, and are $\mathrm{U}(3)$ scalars. On the
other hand, $u^a$ and $v_a$ are $\SO$ scalars and transform under the
$\mathbf{3}$ and $\mathbf{3}^*$ representations of $\mathrm{U}(3)$
\be\label{uvtransformation} \d u^a=u^b(\xi^c_{\ph c d}\s_c^{\ph c
d})_b^{\ph b a}\,,\qquad\quad\d v_a=(\xi^c_{\ph c d}{\s^*}_c^{\ph c
d})_a^{\ph a b}v_b\,,\ee where \be (\s_a^{\ph a b})_c^{\ph c
d}\equiv-i\d_a^{\ph a d}\d_c^{\ph c b}\,.\no\ee Obviously the
hermitian conjugate fields $u_a^*$ and ${v^a}^*$ transform under
$\mathbf{3}^*$ and $\mathbf{3}$ representations of $\mathrm{U}(3)$.

We can naturally define the covariant derivatives \begin{align*}
&\nabla\t^\a=\de \t^\a-\fr,1,2\t^\b(J_{mn}\s^{mn})_\b^{\ph\b\a}\,,\no\\
&\nabla\p_{\ad}=\de\p_{\ad}-\fr,1,2\p_{\bd}(J_{mn}\sb^{mn})^{\bd}_{\ph\bd\ad}\end{align*}
and \begin{align*}
&\nabla u^a=\de u^a+iJ^a_{\ph a b}u^b\,,\\[2ex]
&\nabla v_a=\de v_a-iv_bJ^b_{\ph b a}\,.\end{align*}

The conditions (\ref{uv}) and (\ref{norm}) allow to arrange the
$(u,v)$ variables in the $\mathrm{SU}(3)$ matrix \cite{GIKOS}
\be\label{matrix}
U=\left(u^a\quad\ve^{abc}v_bu_c^*\quad{v^a}^*\right)\,,\no\ee
furthermore they are invariant under $\mathrm{U}(1)$ phase
transformations of $u$ and $v$ separately. Thus the degrees of
freedom of $(u, v)$ are described by the
$\mathrm{SU}(3)/\mathrm{U}(1)_u\!\times\!\mathrm{U}(1)_v$ coset. The
covariant canonical form of the $su(3)$ Lie algebra, $U^{-1}\nabla
U$, projected onto $su(3)\ssm[u(1)\oplus u(1)]$, is \be
j\equiv\begin{pmatrix}0&-{j_1}^*&-{j_2}^*\\j_1&0&-{j_3}^*\\j_2&j_3&0\end{pmatrix}\,,
\no\ee where \be\label{jcurrents} j_1=\ve_{abc}{v^a}^*u^b\nabla
u^c\,,\quad\qquad j_2=v_a\nabla u^a\,,\quad\qquad
j_3=\ve^{abc}u_a^*v_b\nabla v_c\,,\ee and can be used to describe
the $(u, v)$ sector.

The constraint on $\l_3$ admits solution identical to $\l_1$ and,
recalling the one-to-one correspondence between the eigenspaces
$\H_1$ and $\H_3$, we set \be\label{solutionlambda3}
\lab,\a,a=\pb^\a v_a\,,\qquad\qquad \lba,\ad,a=\tb_{\ad} u^a\ee with
the $\mathrm{SO}(3)$ spinors $\pb$ and $\tb$ given by \be
\ph{.}\quad\pb^\a\equiv\p_{\ad}^*(\sb^2)^{\ad\a}\,,\qquad\quad\tb_{\ad}\equiv{\t^\a}^*(\s^2)_{\a\ad}\,.\no\ee

As far as the antighosts are concerned, we introduce a pair of spinors $\o^\a$ and
$\r_{\ad}$ with $(1,0)$ conformal weight that will play the role of
the conjugate momenta of $\t^\a$ and $\p_{\ad}$ respectively. On
general grounds, $w_3$ can be written as \begin{align*}
&\wab,\a,a=\o^\a(u_a^*+A\ve_{abc}u^b{v^c}^*+Bv_a)\,,\\[2ex]
&\wba,\ad,a=\r_{\ad}({v^a}^*+C\ve^{abc}u_b^*v_c+Du^a)\,,\end{align*} where
$A, B, C, D$ are arbitrary functions. Exploiting the gauge
invariance (\ref{anotherlocalinvariance}), which in our notation reads \bea
&&\d\wab,\a,a=\fr,i,2\Omega_m\p_{\ad}(\sb^m)^{\ad\a}v_a+\fr,i,{\sqrt{2}}\ve_{abc}\Omega^b u^c\t^\a \,,\no\\
&&\d\wba,\ad,a=\fr,i,2\Omega_m\t^\a(\s^m)_{\a\ad}u^a-\fr,i,{\sqrt{2}}\ve^{abc}\Omega_bv_c\p_{\ad}\,,\no
\eea we can finally set \be\ph{.}\quad\wab,\a,a=\o^\a
u_a^*\,,\qquad\qquad\wba,\ad,a=\r_{\ad} {v^a}^*\,.\no\ee In
analogous way, $w_1$ can be written as\be\waa,\a,a=\rb^\a
{v^a}^*\,,\qquad\qquad\wbb,\ad,a=\ob_{\ad} u_a^*\no\ee with
\be\ph{.}\quad\rb^\a\equiv\r_{\ad}^*(\sb^2)^{\ad\a}\,,\qquad
\ob_{\ad}\equiv{\o^\a}^*(\s^2)_{\a\ad}\,.\no\ee In conclusion the
pure spinors $\l$ and their conjugate momenta $w$ are given by the
unconstrained $\mathrm{SO}(3,1)$ ghost spinors $\t,\p$ and their conjugate momenta
$\o,\r$ plus the $\textrm{SU}(3)/\textrm{U}(1)\x\textrm{U}(1)$
currents $j_k$. The next step is to construct an action for these variables.

\section{Action with unconstrained ghosts}

We now propose to replace the ghost action $S_\l$ in
(\ref{explicitSlambda}) with an action for the unconstrained ghost
and for the currents $j_k$ and write the $\ads$ superstring action
as \be\label{Stotal} S=S_{matter}+S_{ghost}+S_j\,,\ee where
$S_{matter}$ is given in (\ref{explicitSmatter}),\begin{align}
&S_{ghost}=-\fr,{\ph
,iR^2},{2\pi}\idz\left(\e_{\a\b}\o^\a\onabla\t^\b+\e^{\ad\bd}\r_{\ad}\onabla
                        \p_{\bd}+\e_{\a\b}\rb^\a\nabla\pb^\b+\e^{\ad\bd}\ob_{\ad}\nabla\tb_{\bd}\right)\no\\
&\ph{S_{ghost}=}-\fr,1,{8\pi R^2}\idz
\eta_{m[k}\eta_{l]n}L^{mn}\bar{L}^{kl}\,,\label{Sghostalpha}\end{align}and\begin{align}
&S_j=\fr,R^2,{2\pi}\idz \tr({\oj}^\dag
j)=\fr,R^2,{2\pi}\idz[\sum_{k=1}^3{\oj_k}^*j_k+\mbox{c.c.}]\,,\label{Sj}\end{align}
where the normalization in $S_j$ is chosen for later convenience.
The second line of the action (\ref{Sghostalpha}) gives the coupling
between the $\SO$ ghost currents $L^{mn}$, $\bar{L}^{mn}$ via the
local $\mathrm{AdS}_4$ curvature tensor. These currents can be read
from the ghost coupling to the gauge fields $\oJ_{mn}$ and $J_{mn}$
in (\ref{Sghostalpha}) and are \begin{align*}
&L^{mn}=-\fr,{iR^2},2\left(\o^\a\sbb,m,n,\a,\b\t^\b+\r_{\ad}\saa,m,n,\ad,\bd\p_{\bd}\right)\,,\\
&\bar{L}^{mn}=-\fr,{iR^2},2\left({\rb}^\a\sbb,m,n,\a,\b{\pb}^\b+{\ob}_{\ad}\saa,m,n,\ad,\bd{\tb}_{\bd}\right)\,.\end{align*}
As we will see, the ghost current coupling is necessary to have
one-loop finiteness. Using (\ref{sigmasigma})-(\ref{sigmasigmabar}),
the ghost action (\ref{Sghostalpha}) can be written as
\begin{align}
&S_{ghost}=\fr,R^2,{2\pi}\idz\left[-i\left(\e_{\a\b}\o^\a\onabla\t^\b+\e^{\ad\bd}\r_{\ad}\onabla
                        \p_{\bd}+\e_{\a\b}\rb^\a\nabla\pb^\b+\e^{\ad\bd}\ob_{\ad}\nabla\tb_{\bd}\right)\ph{\fr,1,2}\right.\no\\
&\ph{S_{ghost}=\fr,R^2,{2\pi}\idz\;\,}\left.-\fr,1,2\left(\e_{\a(\g}\e_{\d)\b}\o^\a\t^\b{\rb}^\g{\pb}^\d
               +\e^{\ad(\gd}\e^{\dd)\bd}\r_{\ad}\p_{\bd}{\ob}_{\gd}{\tb}_{\dd}\right)\right]\label{Sghost}\end{align}
with
$\e_{\a(\g}\e_{\d)\b}=-(\e_{\a\g}\e_{\b\d}+\e_{\a\d}\e_{\b\g})$.
\\
\\
By construction $S_{matter}$ and $S_{ghost}$ are invariant under
$\SO\x\mathrm{U}(3)$ local transformations. $S_j$ is obviously
invariant under $\SO$. As far as the $\mathrm{U}(3)$ is concerned,
we recall that $u$ and $v$ transform as \be u^a\:\rightarrow\:u^b
M_b^{\ph b a}\,,\qquad\qquad v_a\:\rightarrow\:M_a^{*\, b}v_b\no\ee
with $M\in\mathrm{U}(3)$ (see (\ref{uvtransformation})). From the
definitions (\ref{jcurrents}), we see that $j_2$ is an
$\mathrm{U}(3)$ scalar, while $j_1$ and $j_3$ are $\mathrm{U}(3)$
pseudo-scalar, i.e. \be
j_1\:\rightarrow\:(\mathrm{det}M)j_1\,,\qquad
j_3\:\rightarrow\:(\mathrm{det}M^*)j_3\,.\no\ee Therefore $S_j$ is
also invariant under local $\mathrm{U}(3)$ transformation.
\\
\\
In the following sections we will compute the central charge and the
effective action using the background field method \cite{abbott} to
treat the action (\ref{Stotal}).

We first discuss the matter sector. In a coset manifold, it is
natural to expand around an element of the group $g=\tg e^{X/R}$
where $\tg$ is in $\OSP$, $X$ is the quantum fluctuation and $R$ is
a scale which counts the order of the perturbative expansion and can
be identified with the radius of $\mathrm{AdS}_4$. The gauge
invariance of the action under $g\rightarrow ge^{\mathfrak{h}}$ with
$\mathfrak{h}\in\sou$ allows to choose
$X=\sum_{i=1}^3X_i\in\osp\!\smallsetminus\![\sou]$. For the
Maurer-Cartan form one gets \be\label{backgroundexpansion}
J=\tJ+\fr,1,R(dX+[\tJ,X])+\fr,1,{2R^2}[dX+[\tJ,X],X]+O\!\left(\fr,1,{R^3}\right)\ee
with $\tJ=\tg^{-1}d\tg$.

Inserting the expansion (\ref{backgroundexpansion}) in
(\ref{Smatter}) one obtains the kinetic term for matter fluctuations
\begin{align}
&S_{XX}=\fr,1,{2\pi}\idz\str\!\left[\fr,1,2\deb
                                       X_2\de X_2+\deb X_1\de X_3\right]\label{kineticmatter}\\
&\ph{S_{XX}}=\fr,1,{2\pi}\idz\!\left[\fr,1,2\deb X^m\de X_m-\deb X^a\de X_a-i\e_{\a\b}\deb\xaa,\a,a\de\xab,\b,a-i\e^{\ad\bd}\deb\xbb,\ad,a\de\xba,\bd,a\right]\,.\no\end{align}
Similarly, from the decomposition $U=\widetilde{U}e^{x/R}$ where
$\widetilde{U}$ is a fixed $\mathrm{SU}(3)$ matrix and $x\in
su(3)\!\smallsetminus\![u(1)\oplus u(1)]$ is the fluctuation, one writes for $j$ \begin{align}
&j=e^{-x/R}\,\tj\,e^{x/R}+e^{-x/R}\de e^{x/R}\no\\
&\ph{j}=\tj+\fr,1,R(\de
x+[\tj,x])+\cdots\,,\label{jexpansion}\end{align} where \be
x=\begin{pmatrix}0&-x_1^*&-x_2^*\\x_1&0&-x_3^*\\x_2&x_3&0\end{pmatrix}\,\no\ee
and $\tj$ is the projection of
$\widetilde{U}^{-1}\nabla\widetilde{U}$ on $su(3)\ssm[u(1)\oplus
u(1)]$. The kinetic term for $x$ is \be\label{kineticuv}
S_{xx}=\fr,1,{\pi}\idz\sum_{k=1}^3\deb x^*_k\de x_k\,.\ee From the
actions (\ref{kineticmatter}) and (\ref{kineticuv}) one computes the
OPE for the fluctuations \be\label{matterOPE}\begin{aligned}
&X^m(z)X^n(w)=-\eta^{mn}\ln|z-w|^2\,,&&X^a(z)X_b(w)=\d_b^{\ph b a}\ln|z-w|^2\,,\\
&\xaa,\a,a(z)\xab,\b,b(w)=-i\e^{\a\b}\d_b^{\ph b
a}\ln|z-w|^2\,,&&\xbb,\ad,a(z)\xba,\bd,b(w)=-i\e_{\ad\bd}\d_a^{\ph a
b}\ln|z-w|^2\,,\end{aligned}\ee \be\label{jOPE}
x^*_k(z)x_l(w)=-\fr,1,2\d_{kl}\ln|z-w|^2\ee and from the action
(\ref{Sghost}) one computes the OPE for the ghost fields
\begin{align}\label{ghostOPE}
&\o^\a(z)\t^\b(w)=\fr,i,{R^2}\e^{\a\b}\fr,1,{z-w}\,,
&&\r_{\ad}(z)\p_{\bd}(w)=\fr,i,{R^2}\e_{\ad\bd}\fr,1,{z-w}\,.\end{align}

\section{Central charge}

The stress-energy tensor of the action is given by \be
T=T_{matter}+T_{ghost}+T_j\,,\no\ee where \begin{align}
&T_{matter}=-R^2\str\!\left(\fr,1,2J_2J_2+J_1J_3\right)\,,\label{mattertensor}\\
&T_{ghost}=iR^2\left(\e_{\a\b}\o^\a\nabla\t^\b+\e^{\ad\bd}\r_{\ad}\nabla\p_{\bd}\right)\,,\label{ghosttensor}\\[1.2ex]
&T_j=-R^2\tr\left(j^\dag j\right)\,.\label{jtensor}\end{align} By
using the expansions (\ref{backgroundexpansion}), (\ref{jexpansion})
and the OPE (\ref{matterOPE})-(\ref{ghostOPE}), one obtains an
expansion in power of $1/R$ of the central charge \be
c=c^{(0)}+\fr,1,{R^2} c^{(2)}+\fr,1,{R^4}
c^{(4)}+\cdots\,.\label{centralcharge}\ee We will study separately
the matter, ghost and
$j$ sectors.\\
\\
Using (\ref{backgroundexpansion}) the matter part of the
stress-energy tensor (\ref{mattertensor}) without background
currents becomes \begin{align*}
&T_{matter}(\tJ=0)=-\str\left[\dst\fr,1,2\de X_2\de X_2+\de X_1\de X_3\right]+O\!\left(\dst\fr,X^4,{R^2}\right)\\
&\qquad=-\dst\fr,1,2\eta_{mn}\de X^m\de X^n+\de X^a\de
X_a+i\e_{\a\b}\de\xaa,\a,a\de\xab,\b,a+i\e^{\ad\bd}\de\xbb,\ad,a\de\xba,\bd,a+O\!\left(\dst\fr,X^4,{R^2}\right)\,.\end{align*}
Notice that a $O\!\left(X^3/R\right)$ contribute is zero, due to the
symmetry properties of the structure constants of the superalgebra.
By means of the OPE (\ref{matterOPE}) we obtain the matter
contribution to the central charge at the zero order \begin{align*}
&c_{bos.\;matter}^{(0)}=4(\mathrm{AdS})+6(\mathbb{C}\mathrm{P})=10\,,&&
c_{ferm.\; matter}^{(0)}=-12-12=-24\,.\end{align*} Moreover, the
absence of a term $1/R$ in $T_{matter}$ implies that
$c^{(2)}_{matter}=0$.
\\
\\
The ghost contribution to the central charge can be computed by
setting $\tJ_{mn}=0$ in (\ref{ghosttensor}) and using the OPE
(\ref{ghostOPE}). At zero order one obtains \be
c_{ghost}^{(0)}=4+4=8\,.\no\ee Moreover corrections to the ghost
central charge arising from the expansion of $J_{mn}$ are
proportional to $1/R^4$, since $X_0=0$.
\\
\\
Finally, using (\ref{jexpansion}), the stress-energy tensor of the
$j$ sector, becomes \be T_j(\tj=0)=-2\sum_{k=1}^3\de x^*_k\de
x_k+O\!\left(\fr,x^4,{R^2}\right)\,,\no\ee and gives \be
c_j^{(0)}=6\,.\no\ee As for the matter sector, $c^{(2)}_j$ is zero
since the $1/R$ term of $T_j$ vanishes.
\\
\\
Collecting the above results one gets \be
c=c_{bos.\,matter}+c_{ferm.\,matter}+c_{ghost}+c_j=10-24+8+6=0\,,\no\ee
up to the $1/{R^2}$ order.

\section{One-loop effective action}

We now discuss the one-loop finiteness of the effective action using
the background field method. In general, by a dimensional analysis,
the one-particle irreducible diagrams which can diverge are the
background field two-point functions, the background
field-ghost-ghost vertices and the four-ghost vertices.

To perform the loop integrals one has to go to momentum space and
use dimensional regularization. Since we are interested in analysing
the UV divergences, we can use the following dictionary
\cite{deboerskenderis,mazzucatovallilo} relating the short distance singularities to $1/\epsilon$ poles: \bea
\ln|0|^2&\rightarrow&-\fr,1,\ve\no\\
\d(z-w)\ln|z-w|^2&\rightarrow&-\fr,1,\ve\label{dictionary}\\
\fr,1,{2\pi}\fr,1,{|z-w|^2}&\rightarrow&\ph{-}\fr,1,\ve\,.\no\eea

\subsection{Background field two-point functions}
The expansion (\ref{backgroundexpansion}) in the  action
(\ref{Stotal}) gives the interactions between the background
fields\footnote{In the following we will omit the \emph{tilde} on
background fields.} and the fluctuations. For the one-loop
background field two-point functions one has to consider only
interactions with two fluctuations and one or two background fields.
Indeed one can easily see that the ghosts do not contribute to these
two-point functions.

It turns out that the results written in terms of supergroup
structure constants are formally analogous in all semi-symmetric
spaces and in particular they are identical to the
$\mathrm{PSU}(2,2|4)$ case \cite{mazzucato,vallilo}. Specifically,
the divergent part of the $\oJ_iJ_i$ two-point functions ($i=1,2,3$)
is proportional to the second Casimir operator  of the supergroup,
that vanishes in $\OSP$ case. Similarly, the divergent contributes
of the one-loop $\oJ_0J_0$ two-point function always sums to zero
due to general properties of the structure constants of a
superalgebra with non-degenerate metric \cite{mazzucato}. As already
mentioned, this result is independent of the ghost sector and
therefore does not provide a test for the solutions
(\ref{solutionlambda1}) and (\ref{solutionlambda3}) of the pure
spinor constraint. On the contrary the one-loop vertices involving
ghost fields depend on the chosen parametrization and therefore
their finiteness is a non-trivial check of the action
(\ref{Stotal}).

\subsection{Background field-ghost-ghost vertices}
We first write the interaction terms of the action (\ref{Stotal})
required at one-loop. In particular, for the $\oJ_0\,\o\t$ vertex,
one needs the interaction term
\begin{align}
&S_{\oJ_0XX}=\fr,1,{4\pi}\idz\!\left\{\oJ_{mn}\left[\de
                                                                   X^m X^n-\de X^n X^m\ph{\fr,1,2}\right.\right.\no\\
&\ph{S_{J_0XX}=1\fr,1,{2\pi}\idz\oJ_{mn}}-\fr,3,4i\left(\de\xab,\a,a\sbb,m,n,\a,\b\xaa,\b,a+
                                                                         \de\xba,\ad,a\saa,m,n,\ad,\bd\xbb,\bd,a\right)\no\\
&\ph{S_{J_0XX}=1\fr,1,{2\pi}\idz\oJ_{mn}}\left.-\fr,1,4i\left(\de\xaa,\a,a\sbb,m,n,\a,\b\xab,\b,a+
                                                                         \de\xbb,\ad,a\saa,m,n,\ad,\bd\xba,\bd,a\right)\right]\no\\
&\ph{S_{J_0XX}=\fr,1,{2\pi}\idz}+\oJ^a_{\ph a b}\left[i\left(\de X^b X_a-\de
                                              X_a X^b\right)-i\d_a^{\ph a b}\left(\de X^c X_c-\de X_c
                                              X^c\right)\ph{\fr,1,2}\right.\no\\
&\ph{S_{J_0XX}=1\fr,1,{2\pi}\idz\oJ_{mn}}+\fr,3,2\left(\e_{\a\b}\de\xab,\a,a\xaa,\b,b-\e^{\ad\bd}\de\xba,\ad,b\xbb,\bd,a\right)\no\\
&\ph{S_{J_0XX}=1\fr,1,{2\pi}\idz\oJ_{mn}}\left.\left.-\fr,1,2\left(\e_{\a\b}\de\xaa,\a,b\xab,\b,a-\e^{\ad\bd}\de\xbb,\ad,a\xba,\bd,b
\right)\right]\right\}\,,\label{J0XX}\end{align} arising from the
matter action (\ref{Smatter}), and the interaction terms \be
S_{XX\o\t}=\fr,1,{8\pi}\idz\left[i\sbb,m,n,\a,\b\deb X_m
X_n-\e_{\a(\g}\e_{\d)\b}(\deb\xaa,\g,a\xab,\d,a+\deb\xab,\g,a\xaa,\d,a)\right]\o^\a\t^\b\,\label{XXot}
\ee and
\begin{align}
&S_{\oJ_0XX\o\t}=-\fr,i,{4\pi}\idz\left[\oJ_{mn}\left(\eta_{kl}\sbb,n,k,\a,\b X^mX^l
+\fr,i,4\e_{\a(\g}\e_{\d)\b}(\s^{kl})_{\eta}^{\ph\eta\g}(\xaa,\eta,a\xab,\d,a+\xab,\eta,a\xaa,\d,a)\right)\right.\no\\
&\left.\ph{S_{\oJ_0XX\o\t}=-\fr,i,{4\pi}\idz}+\oJ^a_{\ph a
b}\e_{\a(\g}\e_{\d)\b}(\xaa,\g,b\xab,\d,a-\xab,\g,a\xaa,\d,b)\right]\o^\a\t^\b\,,\label{J0XXot}\end{align}
arising from the ghost action (\ref{Sghost}).

\newsavebox{\tempbox}
\begin{figure}[h]
\centering \sbox{\tempbox}{\includegraphics[width=6cm]{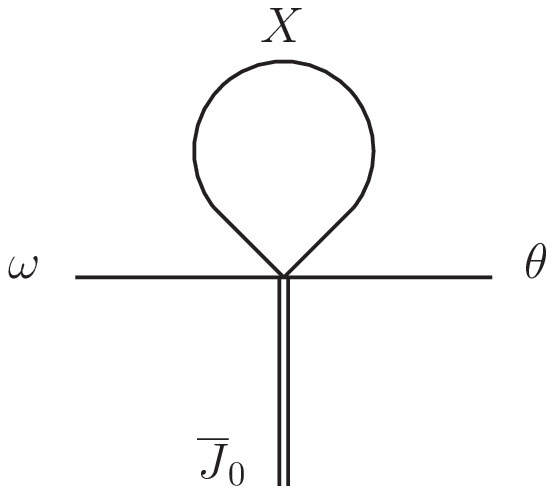}}
\subfloat[]{\usebox{\tempbox}\label{ankhdiagram}} \qquad
\subfloat[]{\vbox to
\ht\tempbox{\vfil\hbox{\includegraphics[width=6cm]{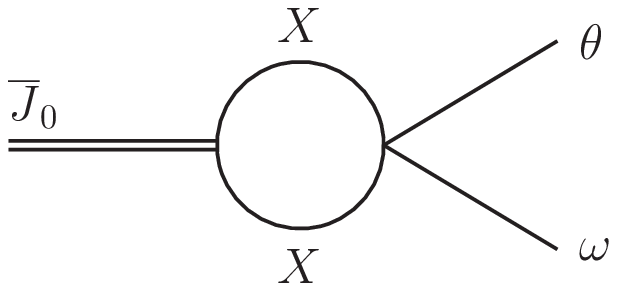}\label{duckdiagram}}\vfil}}
\caption{Background field-ghost-ghost one-loop graphs of the first
(a) and second (b) order.}
\end{figure}

From (\ref{J0XXot}) we get the first order diagram in Figure
\ref{ankhdiagram} which gives the following contribution to the
effective action \be-\fr,i,{2\pi}\idz
\oJ_{mn}(z)\o^\a(z)\sbb,m,n,\a,\b\t^b(z)\ln|0|^2+\mbox{finite
terms}\,.\no\ee From (\ref{J0XX}) and (\ref{XXot}) we get the second
order diagram in Figure \ref{duckdiagram} which gives
\begin{align} &-\fr,i,{4\pi}\idz\idw\oJ_{mn}(z)\o^\a(w)\sbb,m,n,\a,\b\t^\b(w)\no\\
&\ph{\fr,1,{2\pi}\idz\idw}\times\left(-\d(z-w)\ln|z-w|^2+\fr,1,{2\pi}\fr,1,{|z-w|^2}\right)+\mbox{finite
terms}\,.\no\end{align} Summing these contributions and using the
dictionary (\ref{dictionary}), we see that the three-point vertex
$\oJ_0\,\o\t$ is finite at one-loop.

The calculation for the $\oJ_0\,\r\p$ vertex is strictly similar, so
there are no divergencies once again. Analogously the $J_0\,\ob\tb$
and $J_0\,\rb\pb$ functions do not diverge.

\subsection{Four-ghost vertices}

In addition to (\ref{XXot}), the ghost action (\ref{Sghost}) also
gives the interaction \begin{align}
S_{XX\rb\pb}=\fr,1,{8\pi}\idz\left[i\de X_m
X_n\sbb,m,n,\a,\b-\e_{\a(\g}\e_{\d)\b}(\de\xaa,\g,a\xab,\d,a
                                                        +\de\xab,\g,a\xaa,\d,a)\right]{\rb}^\a{\pb}^\b\,.\label{XXrp}
\end{align}
\begin{figure}[h]\centering
\subfloat[][]{\includegraphics[width=5.35cm]{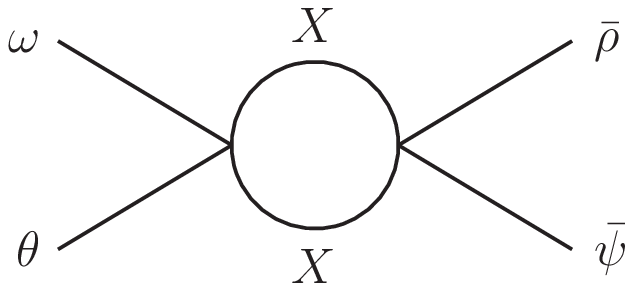}\label{candy_xdiagram}}
\subfloat[][]{\includegraphics[width=5.35cm]{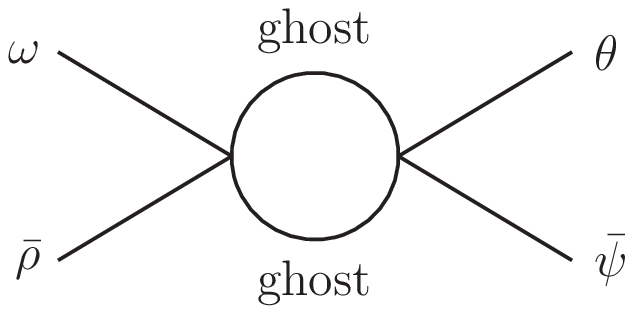}\label{candy_ghostdiagram}}
\subfloat[][]{\includegraphics[width=5.35cm]{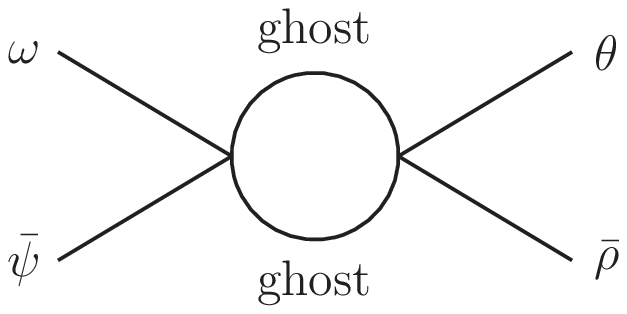}\label{candy_ghost2diagram}}
\caption{Four-ghost graphs with matter loop (a) and ghost loop
(b,c).}
\end{figure}
The interactions (\ref{XXot}) and (\ref{XXrp}) contribute to the
four-ghost function $\o\t\rb\pb$ and yield the matter loop in Figure
\ref{candy_xdiagram}, giving \begin{align}
&-\fr,1,{4\pi}\idz\idw\e_{\a(\g}\e_{\d)\b}\o^\a(z)\t^\b(z)\rb^\g(w)\pb^\d(w)\no\\
&\ph{\fr,1,{2\pi}\idz\idw}\x\left(-\d(z-w)\ln|z-w|^2+\fr,1,{2\pi}\fr,1,{|z-w|^2}\right)+\mbox{finite
terms}\,.\label{candya}\end{align} The remaining diagrams in Figure
\ref{candy_ghostdiagram} and Figure \ref{candy_ghost2diagram}
originate from the four-ghost terms in the action (\ref{Sghost}) and
give \be
-\fr,1,{(2\pi)^2}\idz\idw\left(\e_{\a\g}\e_{\b\d}+\fr,1,4\e_{\a\b}\e_{\g\d}\right)
                                 \o^\a(z)\rb^\g(z)\t^\b(w)\pb^\d(w)\fr,1,{|z-w|^2}+\mbox{finite
terms}\no\ee and \be
-\fr,1,{(2\pi)^2}\idz\idw\left(\e_{\a\d}\e_{\b\g}-\fr,1,4\e_{\a\b}\e_{\g\d}\right)
                                 \o^\a(z)\pb^\d(z)\t^\b(w)\rb^\g(w)\fr,1,{|z-w|^2}+\mbox{finite
terms}\no\ee respectively. Thus for $w\rightarrow z$ the two
diagrams with ghost loop give \be
\fr,1,{2\pi}\idz\idw\e_{\a(\g}\e_{\d)\b}
\o^\a\t^\b\rb^\g\pb^\d\fr,1,{2\pi}\fr,1,{|z-w|^2}+\mbox{finite
terms}\,.\label{candybc}\ee Summing up the contributions
(\ref{candya}) and (\ref{candybc}) one finds that this four-ghost
vertex function is finite. The calculation for the diagrams with
external ghosts $\ob\tb\,\r\p$ is identical.

\section{Discussion and outlook}

In this work we reformulated the ghost sector of the pure spinor
superstring for the $\ads$ superspace in terms of a new set of
unconstrained variables. In the supercoset formulation, the pure
spinor ghosts $\l_1$, $\l_3$ and their conjugate momenta $w_3$,
$w_1$ take values in the fermionic eigenspaces $\H_1$, $\H_3$ of the
$\OSP$ Lie superalgebra and they are subject to the pure spinor
condition $\{\l_1,\l_1\}=\{\l_3,\l_3\}=0$. We replaced these
variables in terms of the (anti)ghosts $(\o,\t)$ and $(\r,\p)$, and
their complex conjugates, which are free $\SO$ Weyl spinors. The
remaining degrees of freedom parametrize a
$\mathrm{SU}(3)/\mathrm{U}(1)\!\times\!\mathrm{U}(1)$ coset.

We wrote an action for these new variables, which is
$\SO\x\mathrm{U}(3)$ gauge invariant. It also contains a coupling
between the ghost Lorentz currents $L^{mn}$ and no other four-ghost
coupling. The model is constructed to have a tree-level vanishing
central charge, since our variables solve the pure spinor
constraint. Using the background field method we showed this also
holds up to the second order in the expansion parameter. Then we
analyzed the possible UV divergences of the effective action and
showed its one-loop finiteness. The ghost Lorentz current
interaction term included in the action is crucial to prove this
fact.

Further checks on our model may be done. In particular, the
understanding of the role of the BRST invariance of the original
pure spinor action. Obviously our model is not BRST invariant. This
also happens in the flat space, where the action written in terms of
the $\mathrm{U}(5)$ variables which solve the pure spinor constraint
is no longer BRST invariant, although it is equivalent to a BRST
invariant gauge-fixed action, modulo a redefinition of the
antighosts \cite{odatonin}. A related and crucial problem would be
the construction of the vertex operators in terms of the new
variables. We plan to address these issues in the nearest future.

Moreover, as already noted, the coset description of the $\ads$
background cannot describe all possible string configurations,
indeed when the string moves entirely in $\mathrm{AdS}_4$, the
fermionic symmetry of the sigma model removes too many degrees of
freedom. This fermionic symmetry is broken in the pure spinor
formulation, thus it would be interesting to see if (and how) this
issue manifests in the present model.

\section*{Acknowledgments}
We thank Dmitri Sorokin for interesting discussions and useful
comments.

\appendix

\section{Conventions}\label{appendixconventions}
For the antisymmetric 2-dimensional tensor  we use the following
convention: \be {\e}^{12}=-{\e}^{21}=1\,, \qquad
{\e}_{12}=-{\e}_{21}=-1\,, \no\ee \be
{\e}_{\a\g}{\e}^{\g\b}={\d}_{\a}^{\ph\a\b}\,,\qquad{\e}^{\ad\gd}{\e}_{\gd\bd}={\d}^{\ad}_{\ph\ad\bd}\,.\no\ee
To treat the symplectic part of the $\OSP$ superalgebra we define
the 4-dimensional charge conjugation matrix ($\m,\n=1,\ldots,4$)\be
C_{\m\n}=\left(\bm
          \e_{\a\b}&0\\
          0&\e^{\ad\bd}
          \enm\right)=
     \left(\ba{c|c}
     \bm
         0&-1\\
         1&0
     \enm&0\\
    \hline
   0&\bm
       0&1\\
       -1&0
      \enm
\ea \right)\no\ee and its inverse \be C^{\m\n}=\left(\bm
          \e^{\a\b}&0\\
          0&\e_{\ad\bd}
          \enm\right)\,,\no\ee so that $C_{\m\r}C^{\r\n}=\d_\m^{\ph\m\n}$, where obviously
\be \d_\m^{\ph\m\n}=\left(\bm
          {\d}_{\a}^{\ph\a\b}&0\\
          0&{\d}^{\ad}_{\ph\ad\bd}
          \enm\right)\,.\no\ee By definition the Dirac matrices $\g^m$ in 4 dimensions
satisfy the Clifford algebra ($m=0,\ldots,3$) \begin{align*}
\{{\g}^m,{\g}^n\}=2{\eta}^{mn}\end{align*}
with ${\eta}^{mn}=\mathrm{diag}(+,-,-,-)$. In the chiral
representation they can be written as \be ({\g}^m)_\m^{\phantom\m\n}=
\left( \bm
0&({\s}^m)_{\a\ad}\\
({\sb}\,^m)^{\ad\a}&0 \enm \right)\no\ee with
${\s}^m=(\mathbf{1},{\s}^1,{\s}^2,{\s}^3)$, being $\s^i$ the Pauli
matrices and
\be({\sb}\,^m)^{\ad\a}={\e}^{\ad\bd}{\e}^{\a\b}({\s}^m)_{\b\bd}\,,\no\ee
i.e. ${\sb}\,^m=(\mathbf{1},-{\s}^1,-{\s}^2,-{\s}^3)$.

\noindent We introduce a $(1+4)$-index $\ms=(0',m)$ and define the
matrices $\g^{\ms\ns}=-\g^{\ns\ms}$  \begin{align*}
&({\g}^{0'm})_\m^{\ph\m\n}\equiv i{\g}^m\,,\\
&({\g}^{mn})_\m^{\ph\m\n}\equiv\displaystyle{\frac{1}{2}}[{\g}^m,{\g}^n]=
\left( \bm
({\s}^{mn})_{\a}^{\ph\a\b}&0\\
0&({\sb}\,^{mn})^{\ad}_{\ph\ad\bd} \enm \right)\end{align*} with
\begin{align*}
&\sba,m,n,\a,\b\equiv\displaystyle{\frac{1}{2}}\left(({\s}^m)_{\a\ad}({\sb}\,^n)^{\ad\b}-({\s}^n)_{\a\ad}({\sb}\,^m)^{\ad\b}\right)\,,\\
&\sab,m,n,\ad,\bd\equiv\displaystyle{\frac{1}{2}}\left(({\sb}\,^m)^{\ad\a}({\s}\,^n)_{\a\bd}-({\sb}\,^n)^{\ad\a}({\s}\,^m)_{\a\bd}\right)\,.
\end{align*} Useful identities are
\begin{align}
&(\s_{mn})_{\a\b}\sbb,m,n,\g,\d=4(\e_{\a\g}\e_{\b\d}+\e_{\a\d}\e_{\b\g})\equiv -4\e_{\a(\g}\e_{\d)\b}\,,\label{sigmasigma}\\
&(\sb_{mn})^{\ad\bd}\saa,m,n,\gd,\dd=4(\e^{\ad\gd}\e^{\bd\dd}+\e^{\ad\dd}\e^{\bd\gd})\equiv -4\e^{\ad(\gd}\e^{\dd)\bd}\,,\\
&(\s_{mn})_{\a\b}\saa,m,n,\ad,\bd=0\,.\label{sigmasigmabar}\end{align}
The indices of $\g^m$ can be raised and lowered by the charge
conjugation matrix \begin{align}\label{gammacharge}
&(\g^m)^{\m\n}\equiv
                    C^{\m\r}(\g^m)_\r^{\ph\r\n}\,,&&(\g^m)_{\m\n}\equiv(\g^m)_\m^{\ph\m\r}C_{\r\n}
\end{align} and the Clifford algebra can be written as
\be\label{cliffordalgebra}
(\g^m)_{\m\r}(\g^n)^{\r\n}+(\g^n)_{\m\r}(\g^m)^{\r\n}=2{\eta}^{mn}\d_{\m}^{\ph\m\n}\,.\no\ee
Notice that $\g^{mn}$ can be also written in term of the
(\ref{gammacharge}) matrices \be\label{gammamn}
(\g^{mn})_{\m}^{\ph\m\n}=\dst\fr,1,2\left((\g^m)_{\m\r}(\g^n)^{\r\n}-(\g^n)_{\m\r}(\g^m)^{\r\n}\right)\,.\ee
The matrix $C$ also raises and lowers the indices of $\g^{\ms\ns}$
\begin{align*}&(\g^{\ms\ns})^{\m\n}\equiv
                    C^{\m\r}(\g^{\ms\ns})_\r^{\ph\r\n}\,,&&(\g^{\ms\ns})_{\m\n}\equiv(\g^{\ms\ns})_\m^{\ph\m\r}C_{\r\n}\,,\end{align*}
explicitly \begin{align*} &({\g}^{0'm})_{\m\n}= i\left( \bm
0&({\s}^m)_{\a\ad}\,{\e}^{\ad\bd}\\
({\sb}\,^m)^{\ad\a}{\e}_{\a\b}&0 \enm \right) \equiv i\left( \bm
0&({\s}^m)_{\a}^{\ph\a\bd}\\
({\sb}\,^m)^{\ad}_{\ph\ad\b}&0 \enm \right)\,,\no\\
&({\g}^{mn})_{\m\n}= \left( \bm
({\s}^{mn})_{\a}^{\ph\a\g}\,{\e}_{\g\b}&0\\
0&({\sb}\,^{mn})^{\ad}_{\ph\ad\gd}\,{\e}^{\gd\bd} \enm \right)
\equiv \left( \bm
\sbb,m,n,\a,\b&0\\
0&\saa,m,n,\ad,\bd \enm \right)\,.\end{align*} Notice that by
definition $C^{-1}\g^mC=-(\g^m)^T$, so $C^{-1}\g^{mn}C=-(\g^{mn})^T$
i.e. $\g^{mn}C=(\g^{mn}C)^T$, using $C^T=-C$. Thus
$(\g^{\ms\ns})_{\m\n}$ and $(\g^{\ms\ns})^{\m\n}$ are symmetric in
$\m,\n$ indices.
\\
\\
To treat the orthogonal part of the superalgebra, we define the
$4\x4$ antisymmetric chiral matrices $\r^M$ ($M=1,\ldots,6$)
satisfying the Clifford algebra \begin{align*}
({\r^M})_{\as\cs}(\r^N)^{\cs\bs}+({\r^N})_{\as\cs}(\r^M)^{\cs\bs}=2{\d}^{MN}{\d}_{\as}^{\ph\as\bs}
\end{align*} with $\as=1,\ldots,4$ and\begin{align}\label{rhomatrices}
(\r^M)^{\as\bs}=\dst{\fr,1,2}{\ve}^{\as\bs\cs\ds}(\r^M)_{\cs\ds}
\qquad\mbox{i.e.}\qquad
(\r^M)_{\as\bs}=\dst{\fr,1,2}{\ve}_{\as\bs\cs\ds}(\r^M)^{\cs\ds}\,,
\end{align}where ${\ve}^{\as\bs\cs\ds}$ is the completely antisymmetric
tensor (${\ve}^{1234}=1$). As in (\ref{gammamn}), we define the
matrices \begin{align*}
(\r^{MN})_{\as}^{\ph\as\bs}\equiv\displaystyle{\frac{1}{2}}\left(({\r}^M)_{\as\cs}(\r^N)^{\cs\bs}-({\r}^N)_{\as\cs}(\r^M)^{\cs\bs}\right)\,.
\end{align*}

\section{$\OSP$ superalgebra}\label{appendixalgebra}

The natural form of the $\mathrm{OSp}(4|6)$ superalgebra is in
$\mathrm{Sp}(4)\times \mathrm{SO}(6)$ basis. Denoting by
$O_{\m\n}=O_{\n\m}$ and $O_{MN}=-O_{NM}$ the bosonic generators of
$\mathrm{Sp}(4)$ and $\mathrm{SO}(6)$ respectively and by $O_{\m M}$
the fermionic ones, the algebra writes \begin{align*}
&[O_{\m\n},O_{\r\s}]=C_{\m\r}O_{\n\s}+C_{\m\s}O_{\n\r}+C_{\n\r}O_{\m\s}+C_{\n\s}O_{\m\r}\\
&[O_{MN},O_{KL}]=\d_{MK}O_{NL}-\d_{ML}O_{NK}-\d_{NK}O_{ML}+\d_{NL}O_{MK}\\
&\{O_{\m M},O_{\r L}\}=i(-\d_{ML}O_{\m\r}+C_{\m\r}O_{ML})\\
&[O_{\m\n},O_{\r L}]=C_{\m\r}O_{\n L}+C_{\n\r}O_{\m L}\\
&[O_{MN},O_{\r L}]=\d_{ML}O_{\r N}-\d_{NL}O_{\r M}\end{align*} with
$\m,\n=1,\ldots,4$ and $M,N=1,\ldots,6$.
\\
\\
Due to the homomorphisms $\mathrm{Sp}(4)\cong\mathrm{SO}(3,2)$ and
$\mathrm{SO}(6)\cong\mathrm{SU}(4)$, the $\mathrm{OSp}(4|6)$
superalgebra can be written in $\mathrm{SO}(3,2)\times
\mathrm{SU}(4)$ basis. The generators of $\mathrm{SO}(3,2)$ and
$\mathrm{SU}(4)$ are obtained by the change of basis
\begin{align*}&
M^{\ms\ns}=\displaystyle{\frac{1}{4}}(\g^{\ms\ns})^{\m\n}O_{\m\n}\,,&&
O_{\m\n}=-\displaystyle{\frac{1}{2}}(\g^{\ms\ns})_{\m\n}M_{\ms\ns}\,,\\[1.5ex]
&U_{\as}^{\ph\as\bs}=-\dst{\fr,i,4}(\r^{MN})_{\as}^{\ph\as\bs}\,O_{MN}\,,&&
O_{MN}=-\dst{\fr,i,2}(\r^{MN})_{\as}^{\ph\as\bs}\,U_{\bs}^{\ph\bs\as}\,.\end{align*}
For the fermionic generators, we define
\begin{align*}&{\cal
O}_{\m\,\as\bs}=\dst{\fr,1,2}O_{\m M}(\r^M)_{\as\bs}\,,&& O_{\m
M}=-\dst{\fr,1,2}(\r^M)^{\as\bs}\,{\cal
O}_{\m\,\as\bs}\,.\end{align*} It is useful to introduce also the
generators
\begin{align}\label{fermionicalternative}&{\cal O}_\m^{\ph\m\as\bs}\equiv\dst{\fr,1,2}{\ve}^{\as\bs\cs\ds}{\cal
O}_{\m\,\cs\ds}=\dst{\fr,1,2}O_{\m M}(\r^M)^{\as\bs}\,,
\end{align}
see (\ref{rhomatrices}). In this basis the $\mathrm{OSp}(4|6)$
superalgebra becomes \bea
&&[M^{\ms\ns},M^{\ks\ls}\,]=\eta^{\ns\ks}M^{\ms\ls}-\eta^{\ms\ks}M^{\ns\ls}-\eta^{\ns\ls}M^{\ms\ks}+\eta^{\ms\ls}M^{\ns\ks}\no\\
&&[U_{\as}^{\ph\as\bs}\,,U_{\cs}^{\ph\cs\ds}\,]=
                                  i\left(\d_{\cs}^{\ph\cs\bs}\,U_{\as}^{\ph\as\ds}-\d_{\as}^{\ph\as\ds}\,U_{\cs}^{\ph\cs\bs}\right)\no\\
&&\left\{{\cal O}_{\m\,\as\bs}\,,{\cal
O}_{\n}^{\ph\m\cs\ds}\,\right\}=\tfrac{i}{4}
\left(\d_{\as}^{\ph\as\ds}\,\d_{\bs}^{\ph\bs\cs}-\d_{\as}^{\ph\as\cs}\,\d_{\bs}^{\ph\bs\ds}\right)
                                                                                               (\g^{\ms\ns})_{\m\n}M_{\ms\ns}\no\\
&&\ph{\left\{{\cal O}_{\m\,\as\bs}\,,{\cal
           O}_{\n}^{\ph\m\cs\ds}\,\right\}=}+\tfrac{1}{2}C_{\m\n}\left(\d_{\as}^{\ph\as\cs}\,U_{\bs}^{\ph\bs\ds}
                      -\d_{\bs}^{\ph\bs\cs}\,U_{\as}^{\ph\as\ds}
                      -\d_{\as}^{\ph\as\ds}\,U_{\bs}^{\ph\bs\cs}
                      +\d_{\bs}^{\ph\bs\ds}\,U_{\as}^{\ph\as\cs}\right)\no\\
&&[M^{\ms\ns},{\cal O}_{\m\,\cs\ds}\,]=-\tfrac{1}{2}(\g^{\ms\ns})_\m^{\ph\m\n}{\cal O}_{\n\,\cs\ds}\no\\
&&[M^{\ms\ns},{\cal O}_{\m}^{\ph\m\cs\ds}\,]=-\tfrac{1}{2}(\g^{\ms\ns})_\m^{\ph\m\n}{\cal O}_{\n}^{\ph\n\cs\ds}\no\\
&&[U_{\as}^{\ph\as\bs}\,,{\cal O}_{\m\,\cs\ds}\,]=
                           {\ph -}i\left(\d_{\cs}^{\ph\cs\bs}\,{\cal O}_{\m\,\as\ds}-\d_{\ds}^{\ph\ds\bs}\,{\cal O}_{\m\,\as\cs}
                                                        -\tfrac{1}{2}\d_{\as}^{\ph\as\bs}\,{\cal
O}_{\m\,\cs\ds}\right)\no\\
&&[U_{\as}^{\ph\as\bs}\,,{\cal O}_{\m}^{\ph\m\cs\ds}\,]=
                    -i\left(\d_{\as}^{\ph\as\cs}\,{\cal O}_{\m}^{\ph\m\bs\ds}-\d_{\as}^{\ph\as\ds}\,{\cal O}_{\m}^{\ph\m\bs\cs}
                                                 -\tfrac{1}{2}\d_{\as}^{\ph\as\bs}\,{\cal O}_{\m\ph{\cs\ds}}^{\ph\m\cs\ds}\right)\no
\eea with $\ms,\ns,\ks,\ls=0',0,\ldots,3$ and
$\as,\bs,\cs,\ds=1,\ldots,4$.
\\
\\
We split the bosonic generators as \begin{align*}&
M^{\ms\ns}=(M^{0'm},M^{mn})\,,&& U_{\as}^{\ph\as\bs}=(U_a^{\ph a
b},U_a^{\ph a 4},U_4^{\ph 4 a})\,,\end{align*} with $a=1,2,3$.
$M^{mn}$ and $U_a^{\ph a b}$ generate the subgroups
$\mathrm{SO}(3,1)\subset\mathrm{SO}(3,2)$ and
$\mathrm{U}(3)\subset\mathrm{SU}(4)$ respectively and we call them
generators of ``rotations''. On the other hand, $M^{0'm}$ and
$U_a^{\ph a 4}$, $U_4^{\ph 4 a}$ lie in the cosets
$\mathrm{SO}(3,2)/\mathrm{SO}(3,1)$ and
$\mathrm{SU}(4)/\mathrm{U}(3)$ respectively and we call them
generators of ``translations''. Thus we define \begin{align*}
&P^m\equiv M^{0'm}\,,&&\vb,a \equiv \dst{\fr,1,{\sqrt{2}}}U_a^{\ph
         a4}\,,&\va,a \equiv \dst{\fr,1,{\sqrt{2}}}U_4^{\ph
         4a}\,.\end{align*}
We split the fermionic generators ${\cal O}_{\m\,\as\bs}$ into
$({\cal O}_{\m\,4a},{\cal O}_{\m\,bc})$ and then substitute ${\cal O}_{\m\,bc}$ with
\begin{align*}{\cal O}_\m^{\ph\m4a}=-\dst{\fr,1,2}\ve^{abc}{\cal
O}_{\m\,bc}\,,\end{align*} using (\ref{fermionicalternative}) and
${\ve}^{4\as\bs\cs}=-{\ve}^{\as\bs\cs4}=-{\ve}^{abc}$. Finally we
split the $\m$ index in the ($\a,\ad$) indices and write
\begin{align*}
&\obb,\a,a\equiv {\cal O}_{\m\,4a}\,,&&\oba,\a,a\equiv {\cal
O}_\m^{\ph\m4a}&&\mbox{when}\quad\m=1,2\\
&\oab,\ad,a\equiv {\cal O}_{\m\,4a}\,,&&\oaa,\ad,a\equiv {\cal
O}_\m^{\ph\m4a}&&\mbox{when}\quad\m=3,4\,\no \end{align*} with
$\a,\ad=1,2$. By means of these definitions and
\begin{align*}\vba,a,b\equiv U_a^{\ph a b}-\d_a^{\ph a b}\,U_{c}^{\ph c
c}\,,\end{align*}one writes the $\mathrm{OSP}(4|6)$ superalgebra in
$\mathrm{SO}(3,1)\x\mathrm{U}(3)\x\mbox{``translations''}$ basis \be
\ba{ll}
\mc{2}{l}{[M^{mn},M^{kl}\,]=\eta^{nk}M^{ml}-\eta^{mk}M^{nl}-\eta^{nl}M^{mk}+\eta^{ml}M^{nk}}\\
\mc{2}{l}{[M^{mn},P^k]=\eta^{nk}P^m-\eta^{mk}P^n}\\
\mc{2}{l}{[P^m,P^n]=-M^{mn}}\\[1.2ex]
\mc{2}{l}{[\vba,a,b,\vba,c,d]=i\left(\d_c^{\ph c
                                                 b}\vba,a,d-\d_a^{\ph a d}\vba,c,b\right)}\\
{}[\vba,a,b,\vb,c]=i\left(\d_c^{\ph c b}\vb,a-\d_a^{\ph a
                                                        b}\vb,c\right)&
                                                                        [\vba,a,b,\va,c]=-i\left(\d_a^{\ph a c}\va,b-\d_a^{\ph a b}\va,c\right)\\
\mc{2}{l}{[\vb,a,\va,b]=\tfrac{i}{2}\left(\vba,a,b-\d_a^{\ph a
                                                               b}\vba,c,c\right)}\\[1.2ex]
\{\obb,\a,a,\obb,\b,b\}=-\tfrac{1}{\sqrt{2}}\e_{\a\b}\ve_{abc}\va,c
                                               & \{\oba,\a,a,\oba,\b,b\}={\ph+}\tfrac{1}{\sqrt{2}}\e_{\a\b}\ve^{abc}\vb,c\\
\{\oaa,\ad,a,\oaa,\bd,b\}={\ph+}\tfrac{1}{\sqrt{2}}\e^{\ad\bd}\ve^{abc}\vb,c
                                               & \{\oab,\ad,a,\oab,\bd,b\}=-\tfrac{1}{\sqrt{2}}\e^{\ad\bd}\ve_{abc}\va,c\\
\{\obb,\a,a,\oaa,\bd,b\}=\tfr,1,2\d_a^{\ph a b}(\s^m)_\a^{\ph \a
                                               \bd}P_m &\{\oba,\a,a,\oab,\bd,b\}=\tfr,1,2\d_b^{\ph b a}(\overline{\s}^m)^{\bd}_{\ph \ad\a}P_m\\[1.2ex]
\mc{2}{l}{\{\obb,\a,a,\oba,\b,b\}=-\tfrac{i}{4}\d_a^{\ph a b}\sbb,m,n,\a,\b M_{mn}+\tfrac{1}{2}\e_{\a\b}\vba,a,b}\\
\mc{2}{l}{\{\oab,\ad,a,\oaa,\bd,b\}=-\tfrac{i}{4}\d_a^{\ph a b}\saa,m,n,\ad,\bd M_{mn}+\tfrac{1}{2}\e^{\ad\bd}\vba,a,b}\\
\{\obb,\a,a,\oab,\bd,b\}=0 & \{\oba,\a,a,\oaa,\bd,b\}=0\\[1.2ex]
{}[M^{mn},\obb,\a,a]=-\tfrac{1}{2}\sba,m,n,\a,\b\obb,\b,a & [M^{mn},\oba,\a,a]=-\tfrac{1}{2}\sba,m,n,\a,\b\oba,\b,a\\
{}[M^{mn},\oaa,\ad,a]=-\tfrac{1}{2}\sab,m,n,\ad,\bd\oaa,\bd,a &
                                                                [M^{mn},\oab,\ad,a]=-\tfrac{1}{2}\sab,m,n,\ad,\bd\oab,\bd,a\\[1.2ex]
{}[P^m,\obb,\a,a]=-\tfrac{i}{2}(\s^m)_{\a\bd}\oab,\bd,a & [P^m,\oba,\a,a]=-\tfrac{i}{2}(\s^m)_{\a\bd}\oaa,\bd,a\\
{}[P^m,\oaa,\ad,a]=-\tfrac{i}{2}(\sb^m)^{\ad\b}\oba,\b,a &
                                                          [P^m,\oab,\ad,a]=-\tfrac{i}{2}(\sb^m)^{\ad\b}\obb,\b,a\\[1.2ex]
{}[\vba,a,b,\obb,\a,c]={\ph+}i\d_c^{\ph c b}\obb,\a,a & [\vba,a,b,\oab,\ad,c]={\ph+}i\d_c^{\ph c b}\oab,\ad,a\\
{}[\vba,a,b,\oaa,\ad,c]=-i\d_a^{\ph a c}\oaa,\ad,b & [\vba,a,b,\oba,\a,c]=-i\d_a^{\ph a
                                                                                        c}\oba,\a,b\\[1.2ex]
{}[\vb,a,\obb,\a,b]=-\tfrac{i}{\sqrt 2}\ve_{abc}\oba,\a,c & [\vb,a,\oba,\a,b]=0\\
{}[\vb,a,\oaa,\ad,b]=0 & [\vb,a,\oab,\ad,b]=-\tfrac{i}{\sqrt2}\ve_{abc}\oaa,\ad,c\\
{}[\va,a,\obb,\a,b]=0 & [\va,a,\oba,\a,b]={\ph i}\tfrac{i}{\sqrt2}\ve^{abc}\obb,\a,c\\
{}[\va,a,\oaa,\ad,b]={\ph i}\tfrac{i}{\sqrt 2}\ve^{abc}\oab,\ad,c &
                                                                  [\va,a,\oab,\ad,b]=0\,.\ea\no\ee

\end{document}